\documentclass{aastex631}

\begin{document}

\title{Forecasting SEP Events During Solar Cycles 23 and 24 Using Interpretable Machine Learning}

\author[0000-0002-0972-8642]{Spiridon Kasapis}
\affiliation{NASA Ames Research Center, Moffett Field, CA, USA}

\author[0000-0003-4144-2270]{Irina N. Kitiashvili}
\affiliation{NASA Ames Research Center, Moffett Field, CA, USA}

\author{Paul Kosovich}
\affiliation{New Jersey Institute of Technology, Newark, NJ, USA}

\author[0000-0003-0364-4883]{Alexander G. Kosovichev}
\affiliation{New Jersey Institute of Technology, Newark, NJ, USA}
\affiliation{NASA Ames Research Center, Moffett Field, CA, USA}

\author[0000-0002-4001-1295]{Viacheslav M. Sadykov}
\affiliation{Georgia State University, Atlanta, GA, USA}

\author{Patrick O'Keefe}
\affiliation{New Jersey Institute of Technology, Newark, NJ, USA}

\author{Vincent Wang}
\affiliation{Purdue University, West Lafayette, IN, USA}

\begin{abstract}
Prediction of the Solar Energetic Particle (SEP) events garner increasing interest as space missions extend beyond Earth's protective magnetosphere. These events, which are, in most cases, products of magnetic reconnection-driven processes during solar flares or fast coronal-mass-ejection-driven shock waves, pose significant radiation hazards to aviation, space-based electronics, and particularly, space exploration. In this work, we utilize the recently developed dataset that combines the Solar Dynamics Observatory/Helioseismic and Magnetic Imager's (SDO/HMI) Space weather HMI Active Region Patches (SHARP) and the Solar and Heliospheric Observatory/Michelson Doppler Imager's (SoHO/MDI) Space Weather MDI Active Region Patches (SMARP). We employ a suite of machine learning strategies, including Support Vector Machines (SVM) and regression models, to evaluate the predictive potential of this new data product for a forecast of post-solar flare SEP events. Our study indicates that despite the augmented volume of data, the prediction accuracy reaches $0.7 \pm 0.1$, which aligns with but does not exceed these published benchmarks. A linear SVM model with training and testing configurations that mimic an operational setting (positive-negative imbalance) reveals a slight increase ($+ 0.04 \pm 0.05$) in the accuracy of a 14-hour SEP forecast compared to previous studies. This outcome emphasizes the imperative for more sophisticated, physics-informed models to better understand the underlying processes leading to SEP events.
\end{abstract}


%%%%%%% NEW SECTION %%%%%%% 

\section{Introduction} \label{sec:Intro}

Solar Energetic Particle (SEP) events are one of the manifestations of solar activity that may significantly impact the conditions of the space environment. For example, the large solar particle event of September 2017 emphasized a significant surge in the charged and neutral particle flux that was able to reach Mars' surface \citep{zeitlin2018analysis}. While the doses from this specific event were below NASA's stipulated radiation exposure limits for astronauts, the risk for future explorers is evident. This concern becomes particularly relevant in scenarios where human explorers might be far from their habitats on other celestial bodies, with the onset of an event leaving them vulnerable to enhanced radiation doses. Therefore, forecasting and predicting SEP events is paramount.

SEP events vary in intensity, spanning from suprathermal (few keV) up to relativistic (few GeV) energies, and are accelerated near the Sun either by magnetic reconnection-driven processes during solar flares or by fast Coronal Mass Ejections (CME). SEP events are categorized as impulsive and gradual \citep{desai2016large}. The large gradual SEP events are of special interest because they are accompanied by high-energy protons ($\geq 10$ MeV) that pose serious radiation threats to aviation and space operations. While these events are intriguing from a scientific viewpoint, they have not always been consistently detected. Therefore, while over 200 major solar proton events were observed over the last 40 years, discerning a pattern between these occurrences and the solar cycle remains elusive, adding to the difficulty in inferring the number and intensity of major solar proton events that can affect Earth \citep{shea1990summary}.

Many models have been developed to predict SEP events  \citep{whitman2022review}. These models vary in their intricacies, ranging from physics-based \citep[e.g.,][]{zhang2017precipitation} to empirical \citep[e.g.,][]{anastasiadis2017predicting,georgoulis2021flare} and probabilistic \citep[e.g.,][]{papaioannou2022probabilistic}. Lately, the embrace of machine learning and hybrid model approaches signals a multidimensional approach toward tackling this issue. For instance, the Space Radiation Intelligence System (SPRINTS) merges pre-eruptive and post-eruptive metadata, offering predictions for various solar-driven events, including SEPs \citep{engell2017sprints}.

Leveraging vast datasets, ML models are being trained to predict Solar Proton Events (SPEs) by considering diverse parameters, such as the magnetic field characteristics of solar Active Regions (ARs), the preceding soft X-ray and proton fluxes, and the statistics of solar radio bursts. For example, ML models have demonstrated predictive capabilities superior to conventional SWPC NOAA forecasts, indicating the possibility of developing robust ``all-clear" SPE forecasts \citep[e.g.,][]{sadykov2021prediction}. Similarly, \cite{ji2020all} has demonstrated that through data collection and intricate predictive model building, a time series classification apparatus can offer more precise forecasts compared to baseline models by tuning model hyperparameters.

In this study, we build upon the current state of SEP modeling, utilizing a dataset that spans two solar cycles. This extended dataset offers a new perspective on the SEP prediction problem, aiming to improve the accuracy and reliability of the existing models. To do this, we take advantage of our previous developments \citep{kasapis2022interpretable} and a new dataset that covers two solar cycles, 23 and 24 \citep{kosovich2024time}, briefly described in Section \ref{sec:SHMARP}. Because the prediction method is based on the known connection between SEP events and solar flares \citep{le2017dependence}, Section \ref{sec:Matching} outlines the process of matching these space weather phenomena that follow up statistical analysis presented in Section \ref{sec:Statistics}. Finally, in Sections \ref{sec:ML} and \ref{sec:Results}, we present the results of this study and the ML models, while in Section \ref{sec:Conclusion} we summarize the results that span the two Solar Cycle observations.

%%%%%%% NEW SECTION %%%%%%%  

\section{Solar Cycles 23 and 24 active regions dataset}\label{sec:SHMARP}

It is well known that not all active regions exhibit eruptive activity, and only a few of the ones that do are a source of SEP events \citep{cane2010study}. Over Solar Cycles 23 and 24, although the number of active regions and flares that erupted within them are on the order of thousands, only 168 SEPs were recorded by the NASA Space Radiation Analysis Group\footnote{\url{https://srag.jsc.nasa.gov/}} (SRAG). This creates a significant statistical disproportion expectation of SEP events (so-called class-imbalance problem) when the forecast is based on the thousands of flares, and as a result, models often tend to miss them. Therefore, increasing the size of the dataset suitable for ML training may improve the robustness of the SEP events forecast. Because most SEP events are associated with flares that erupted in ARs, it is natural to utilize the available Space-Weather MDI Active Region Patches \citep[SMARP,][]{bobra2017space} and Space-Weather HMI Active Region Patches \citep[SHARP,][]{bobra2011sharp}. The SMARP and SHARP datasets include maps of automatically-tracked active regions extracted from full-disk magnetograms. These active region patches and their parameters have been used for a number of solar flare prediction studies \citep[e.g.,][]{2015ApJ...798..135B,Kontogiannis2017,Wang2023}. The vast majority of previous studies did not take advantage of both SMARP and SHARP. Training a machine learning model with the SMARP dataset demonstrated a potential to predict SEP events \citep{kasapis2022interpretable}, where training was based on five predictors: the total unsigned magnetic flux, the vertical field gradient, the unsigned flux {\it R} near polarity inversion lines \citep{2007ApJ...655L.117S}, the angular distance between an active region and Earth's magnetic foot point \citep{ippolito2005magnetic}, and the area of an active region. 

We use time series of summary parameters (keywords) of solar active regions from both the SMARP and SHARP data series \citep{bobra2011sharp,bobra2017space,bobra2021smarps}, which were combined into a single consistent dataset \citep{kosovich2024time}. This merged dataset keeps the original cadence of the data products: 96 minutes for MDI and 12 minutes for HMI observations. The SMARP-SHARP dataset includes a continuous set of 21 keywords (Table~\ref{tab:SHMARPKeywords}), representing homogeneous observations of active region patches from SoHO/MDI and SDO/HMI for the period between April 4, 1996, and May 9, 2023. Because MDI did not observe the transverse magnetic field, we computed the vertical components of the unsigned flux and the mean magnetic field gradient from the line-of-sight parameters assuming that the magnetic field of active regions is predominantly radial. For both SHARP and SMARP data sets, the R\_VALUE was recomputed by finding the common antilogarithm of the original R\_VALUE, and SMARP MEANGBL units were converted from Gauss/pixel to Gauss/Mm. Filtering was applied to exclude low-quality observables and any records corresponding to the Stonyhurst coordinates beyond $\pm 65$ degrees longitude to mitigate the foreshortening effect near the solar limb. Six SMARP parameters (USFLUXL, MEANGBL, R\_VALUE, CMASKL, MEANGBZ, and USFLUXZ) were rescaled by applying total least squares fitting, and the two data sets were merged on May 1, 2010, 00:00:00 TAI. For a more thorough overview of filtering, rescaling and merging of the SHARP and SMARP keywords, we  refer readers to \citet{kosovich2024time}. 

\begin{table}[h]
\centering
\begin{tabular}{|l|l|l|}
\hline
\textbf{Keyword} & \textbf{Description} & \textbf{Example} \\ \hline
DBINDEX & Database Index & 'mdi.smarp\_cea\_96m...' \\ \hline
T\_OBS & Time of Observation & 11/4/97 4:47 \\ \hline
UNIX\_TIME & T\_OBS expressed as time elapsed since Unix epoch (days) & 10169.19965 \\ \hline
ARPNUM & The number of an active region patch& 581 \\ \hline
NOAA\_AR & NOAA active region number that best matches ARPNUM & 8100 \\ \hline
NOAA\_ARS & List of all NOAA active regions matching this ARPNUM & 8100 \\ \hline
\textbf{USFLUXL} & \textbf{Total line-of-sight unsigned flux (Maxwells)} & \textbf{3.18E+22} \\ \hline
\textbf{R\_VALUE} & \textbf{Unsigned flux R near polarity inversion lines (Maxwells)} & \textbf{251995.71} \\ \hline
\textbf{MEANGBL} & \textbf{Mean value of line-of-sight field gradient (Gauss/Mm)} & \textbf{103.10} \\ \hline
\textbf{USFLUXZ} & \textbf{Vertical component of the total unsigned flux (Maxwells)} & \textbf{120.44} \\ \hline
\textbf{MEANGBZ} & \textbf{Mean value of the vertical field gradient (Gauss/Mm)} & \textbf{6.75E+22} \\ \hline
LAT\_FWT & Stonyhurst latitude of flux-weighted center of active pixels (degrees) & -19.58 \\ \hline
\textbf{CRLT\_OBS} & \textbf{Carrington latitude of the observer (degrees)} & \textbf{4.02} \\ \hline
LON\_FWT & Stonyhurst longitude of flux-weighted center of active pixels (degrees) & 27.23 \\ \hline
\textbf{CRLN\_OBS} & \textbf{Carrington longitude of the observer (degrees)} & \textbf{324.65} \\ \hline
CAR\_ROT & Carrington rotation number of CRLN\_OBS & 1929 \\ \hline
CMASKL & CEA (cylindrical equal-area) pixels in the active region & 182848.46 \\ \hline
CDELT1 & Map scale in degrees per pixel & 0.12 \\ \hline
DSUN\_OBS & Distance from SOHO/SDO to center of the Sun (meters) & 146808723630.28 \\ \hline
RSUN\_OBS & Observed angular radius of the Sun (arcseconds) & 977.876787 \\ \hline
QUALITY & Quality Index & 512 \\ \hline
\end{tabular}
\caption{List of the merged SMARP-SHARP keywords \citep{kosovich2024time}. In bold are the keywords used as predictors in this study, and as examples, the values that describe AR8100 on 4 November 1997, 04:47:00, are given. The ARPNUM is called ``TARPNUM'' in the SMARP dataset and ``HARPNUM'' in the SHARP dataset, but were renamed here for consistency.}
\label{tab:SHMARPKeywords}
\end{table}

To evaluate the predictive capabilities of an ML-driven SEP analysis, we use parameters of the flaring active regions as precursors, which are stored as the SMARP-SHARP parameters (Table~\ref{tab:SHMARPKeywords}). In our research, we use eight of these parameters (marked bold in Table \ref{tab:SHMARPKeywords}). Five physical parameters (R\_VALUE, MEANGBL, MEANGBZ, USFLUXL, USFLUXZ) are used for the models’ training and evaluation. These keywords (so-called `predictors' in machine learning) are used as input to an ML model. Spatial parameters CRLT\_OBS and CRLN\_OBS are also involved in training and evaluation as they define the magnetic connectivity of active regions to the Earth \citep{ippolito2005magnetic}, which we also use as a predictor. The time variable, T\_OBS, allows us to pick the correct SMARP-SHARP data records associated with a SEP.

%%%%%%% NEW SECTION %%%%%%% 

\section{Coupling of SEP Events and Solar Flares} \label{sec:Matching}

Most SEP events are associated with solar eruptive activity, therefore the NOAA solar X-ray flare Catalog \footnote{\url{https://www.ngdc.noaa.gov/stp/solar/solarflares.html}} and the list of SEP events provided by the NASA SRAG are used to identify the SMARP-SHARP data points (selected from the AR time series) that will train our ML models. The flare catalog contains 4238 flares recorded from April 22, 1996, to December 30, 2022. Only flares of the C, M, and X classes are kept because the weaker flares do not produce SEP events, allowing us to mitigate the class-imbalance problem. \cite{kasapis2022interpretable} showed that predicting whether A and B class flares produced a SEP is a trivial problem for an ML model. Thus, the total number of flares was reduced to 3421. The SEP list includes 168 events recorded from November 4, 1997, to May 9, 2023. We excluded 14 SEP events because they occurred on the farside of the Sun or they were not associated with flares. Our analysis does not include any SEP events on the limb or farside because of the absence of reliable observations. The SEP list provided by NASA SRAG contains 154 SEP-flare couples (SEPs associated with flares). The NOAA list includes only 115 flares that can be verified (matched) with these SEP-flare couples. 

Similar to \cite{kasapis2022interpretable}, we define flares associated with SEP as a \textit{positive} event and the flares that did not produce a SEP event as a \textit{negative} event. Due to data gaps that exist in the merged SMARP-SHARP dataset, an additional 5 positive and 60 negative of these SEP-Flare couples had to be excluded. Therefore, the final count for the flares that will be used for SEP prediction is 110 positive and 3246 negative events. The SMARP-SHARP data points availability and selection are discussed in Section \ref{sec:Statistics}.

\begin{table}[h]
\centering
\begin{tabular}{|l|l|l|} \hline
\multicolumn{3}{|c|}{\textbf{Default NOAA Flare Keywords}} \\ \hline
\textbf{Keyword} & \textbf{Description} & \textbf{Example} \\ \hline
\textit{t\_start} & Flare start time & 1997-11-04 05:52:00 \\ \hline
\textit{t\_max} & Flare maximum intensity time & 1997-11-04 05:58:00 \\ \hline
\textit{t\_end} & Flare end time & 1997-11-04 06:02:00 \\ \hline
\textit{class} & Flare class & X2.1 \\ \hline
\textit{location} & Flare location (NESW format) & S14W33 \\ \hline
\textit{AR} & NOAA Active region number associated with flare & 8100 \\ \hline
\multicolumn{3}{|c|}{\textbf{New Keywords}} \\ \hline
\textbf{Keyword} & \textbf{Description} & \textbf{Example} \\ \hline
\textit{SEP\_Match} & SEP producing flare (True or False) & True \\ \hline
\textit{intensity} & Intensity ($\text{W/m}^2$)& 0.00021 \\ \hline
\textit{coords} & Flare heliographic coordinates (degrees) & (-14, 33) \\ \hline
\textit{ang\_dist} & Angular distance from the magnetic footpoint of the Earth (radians) & -0.32045 \\ \hline
\end{tabular}
\caption{List of keywords in the NOAA flare dataset (top) and the ones created for the convenience of our analysis (bottom). Although the \textit{SEP\_Match} keyword is used for the ML models training, the \textit{intensity}, \textit{coords} and \textit{ang\_dist} keywords are only used for demonstrating the relationship between flares and SEPs in Figure~\ref{fig:FlareKeywordsHistograms}.}
\label{tab:FlareKeywords}
\end{table}

To perform the association of SEP events to solar flares, we use the NOAA flare keywords (\textit{t\_start}, \textit{t\_max}, \textit{t\_end}, \textit{class}, \textit{location}, and \textit{AR}) in Table~\ref{tab:FlareKeywords}. Additional keywords were generated to reflect if a particular flare produced a SEP event or not  (\textit{SEP\_Match} keyword), as well as additional characteristics of an eruptive event (\textit{intensity}, \textit{coords}, and \textit{ang\_dist} keywords). The \textit{SEP\_Match} keyword is particularly important in this project, as it is used as the target for our ML-ready datasets. This SEP-flare matching process has assigned 110 flares as positive (\textit{SEP\_Match} = `True') and 3246 as negative (\textit{SEP\_Match} = `False').

The flare \textit{intensity} keyword in Table~\ref{tab:FlareKeywords} is defined as $I = n \times f$ where $n$ corresponds to the intensity within the flare class (\textit{class} keyword), and $f$ is the classification factor which depends on the class of a flare: $10^{-6}$ for `C' class flares, $10^{-5}$ for `M', and $10^{-4}$ for `X' class flares. The \textit{coords} keyword is produced by converting the solar flare location coordinates from the NSEW format (\textit{location}) to heliographic coordinates. The NSEW format is a string-based representation, where the first character indicates North (N) or South (S), followed by two digits representing the latitude. The fourth character represents East (E) or West (W), followed by two digits for the longitude.  The heliographic coordinates are based on the Sun's rotation and are expressed as latitude and longitude in degrees, with North and West considered positive. Therefore, the NSEW coordinates `S14W33' would be converted to (-14, 33) in heliographic coordinates (Table~\ref{tab:FlareKeywords}, New Keywords). Finally, the \textit{ang\_dist} keyword represents the angular distance $\Delta \sigma$ between the active region and the magnetic footpoint of Earth using the haversine formula \citep{robusto1957cosine}, which is appropriate for calculating great-circle distances between two points on a sphere:
\begin{equation}
\Delta \sigma = \arccos(\sin \theta_1 \sin \theta_2 + \cos \theta_1 \cos \theta_2 \cos(\phi_1 - \phi_2))
\label{eq:AngularDistance}
\end{equation}
where $(\theta_1, \phi_1)$ are the latitude and longitude of an active region, and $(\theta_2, \phi_2)$ are the latitude and longitude of the magnetic footpoint of Earth. The distance to the magnetic footpoint of the Earth is critical for SEP prediction because it delineates the magnetic connectivity between solar flares and the Earth, serving as a pathway for SEPs to travel to our planet's vicinity \citep{ippolito2005magnetic}. We assume that the magnetic footpoint of Earth is at the position $(0, 45)$ degrees in the Stonehurst heliographic coordinates, the location of the magnetic field line connecting the Sun and Earth on the source surface (2.5 solar radii from the solar center) where Parker's spiral originates. The coordinates are initially in degrees, but we convert them into radians to fit the Python trigonometric functions requirement. After the calculation, the angular distance $\Delta \sigma$ is returned in radians. An additional consideration is made for the active regions that are East of the magnetic footpoint, in which case the distance is assigned a negative sign. Thus, using the SHARP-SMARP coordinates (CRLT\_OBS and CRLN\_OBS, Table \ref{tab:SHMARPKeywords}) in Equation~\ref{eq:AngularDistance}, we obtained the angular distance to the magnetic footpoint of the Earth (ANG\_DIST keyword) for the ARs in the SHARP-SMARP dataset.

\begin{figure}[ht]
\centering
\includegraphics[width=\textwidth]{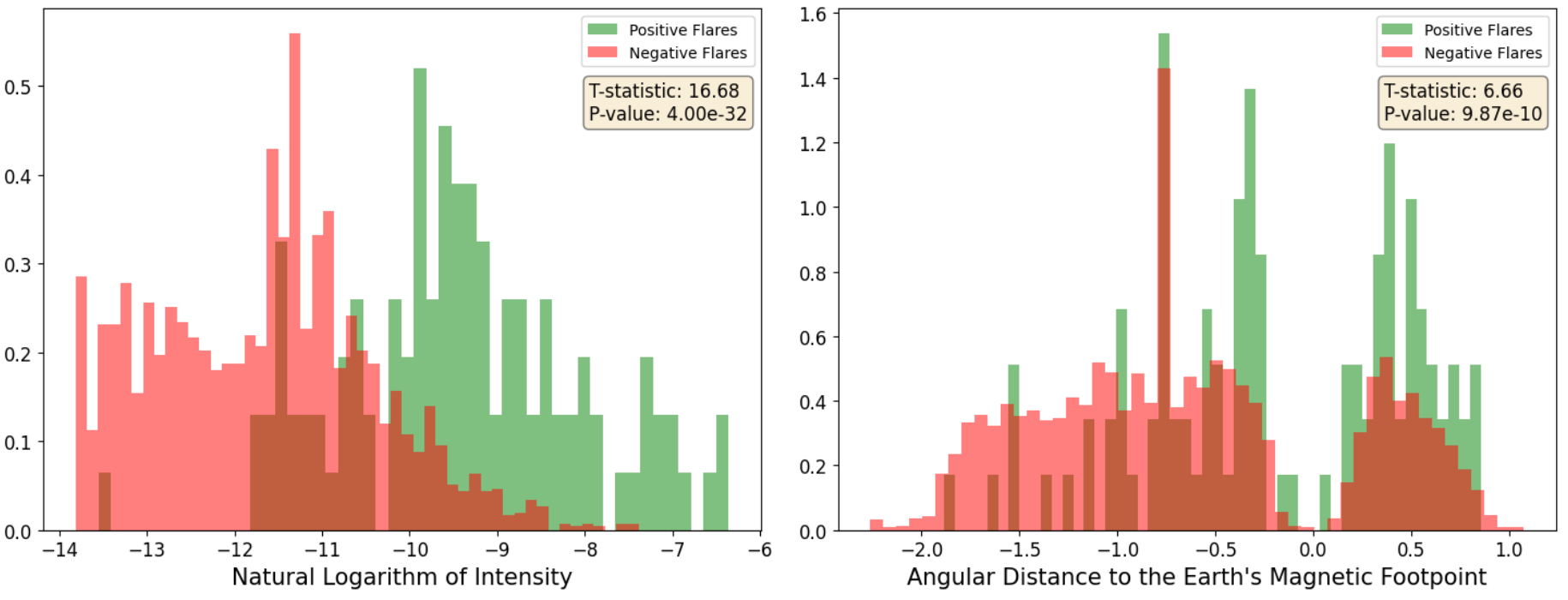}
\caption{Probability density distribution of the 3356 positive (green) and negative (red) events in our dataset that occurred between April 4, 1996, and May 9, 2023 (SHARP-SMARP dataset availability) for the logarithm of flare intensities (left panel) and angular distances to the magnetic footpoint of the Earth (right panel).}
\label{fig:FlareKeywordsHistograms}
\end{figure}

The \textit{ang\_dist} flare keyword is calculated using the flare location in Table~\ref{tab:FlareKeywords}, whereas, in our ML models, we will be using the AR location information from Table~\ref{tab:SHMARPKeywords}. The keyword values in Table~\ref{tab:SHMARPKeywords} are the ones provided in the SMARP-SHARP dataset, and we are going to be evaluating them in this research rather than the NOAA flare information. The solar flare keywords are not used to train the ML models as in an operational setting, at the moment of prediction, we do not have knowledge of the flare eruption and the physical parameters associated with it. The flare keywords are only used to assign targets (SEP or not) to the SMARP-SHARP data points and to shed light on the relationship between flares and the SEPs they produced. For example, the density histogram on the left panel of Figure \ref{fig:FlareKeywordsHistograms} shows much less overlap between the positive (green) and negative (red) flares compared to the histograms on the right panel, indicating higher predictive capabilities of magnetic flux. 

This difference in predictive capability between flare intensity and flare angular distance to the magnetic footpoint of the Earth is also reflected in the values of the t-statistic and the p-value, indicating that if flare intensity is known, then predicting whether a flare will produce a SEP is much easier than only performing prediction based on the flare coordinates. The t-statistic \citep{kim2015t} is an indicator of the difference between two group means in units of standard error; hence, a larger absolute t-statistic signifies a more noticeable difference. The p-value \citep{thisted1998p}, on the other hand, provides a measure of evidence against a null hypothesis. A smaller p-value indicates stronger evidence against the null hypothesis, which, in this context, suggests that the two groups (positive and negative flares) have distinct means.

Although Figure \ref{fig:FlareKeywordsHistograms} demonstrates that \textit{intensity} is a relatively reliable predictor of SEP events, the right panel shows that the angular distance is not, but still carries some information related to SEP production. As expected, flares with low values of angular distance (flares that occurred close to the Earth-Sun magnetic connection point) are more likely to produce SEPs that can be observed at Earth, compared to flares that erupted far away from the magnetic footpoint (Figure \ref{fig:FlareKeywordsHistograms}, right panel, angular distance values $\leq-1$). It is important to note, that such keywords can be used for SEP prediction only if a flare prediction model is able to infer the flare intensity and position on the solar disc \citep{jiao2020solar, chen2021flare}.

%%%% UP TO HERE WITH IRINA %%%%%%%

%%%%%%% NEW SECTION %%%%%%% 

\section{SHARP-SMARP Data Selection and creation of ML-Ready Datasets} \label{sec:Statistics}

\begin{figure}[ht]
\centering
\includegraphics[width=0.9\textwidth]{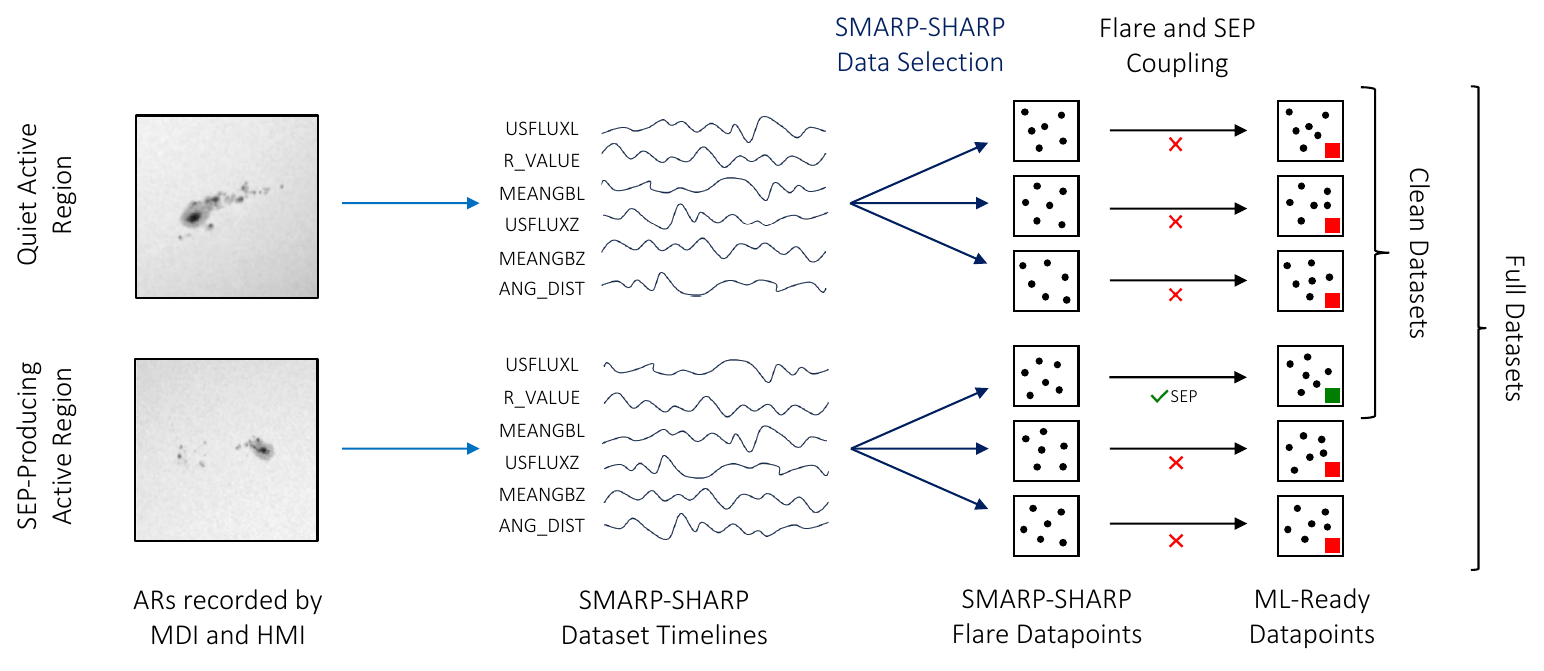}
\caption{Scheme of a workflow to prepare ML-ready datasets from the SMARP and SHARP \citep{bobra2021smarps,kosovich2024time}. The arrows depict the previous work (light black, Section~\ref{sec:SHMARP}), the flare and SEP coupling process (black, Section~\ref{sec:Matching}) and the SMARP-SHARP data selection (blue, Section~\ref{sec:Statistics}).
}
\label{fig:Schematic}
\end{figure}

To predict the SEP events, we consider the time evolution of an active region and its properties during the start time of a corresponding solar flare. In this work, we use 3869 active regions (ARs) from the SMARP-SHARP dataset \citep{kosovich2024time}, 872 of which produced a flare. Because the eruptive activity of the Sun is the primary source of the SEP events, we consider only flaring active regions (such as the two example ARs in Figure~\ref{fig:Schematic}). Thus, we combine information available from the three data sources: 1) properties of flaring active regions from the SHARP-SMARP data set, 2) the NOAA catalog of solar flares, and 3) the SWPC catalog of SEP events.

After labeling every flare in the NOAA flare list (using the SEP\_Match keyword) based on whether it is associated with a SEP (Section~\ref{sec:Matching}), we proceed with selecting the appropriate SHARP-SMARP data (Section~\ref{sec:Statistics}) used to train various ML-models. Figure \ref{fig:Schematic} illustrates these two distinct processes of creating different datasets used in this paper (Flare and SEP Coupling, SMARP-SHARP Data Selection). 

%Section~\ref{sec:Matching}  describes the process of labeling every flare in the NOAA flare list (using the SEP\_Match keyword) based on whether it is associated with a SEP. Given this knowledge, in this Section we will describe the process of selecting the appropriate SHARP-SMARP Data used to train various ML-models. Figure \ref{fig:Schematic} illustrates these two distinct processes of creating different datasets used in this paper (Flare and SEP Coupling, SMARP-SHARP Data Selection).  

\begin{figure}[ht]
\centering
\includegraphics[width=0.9\textwidth]{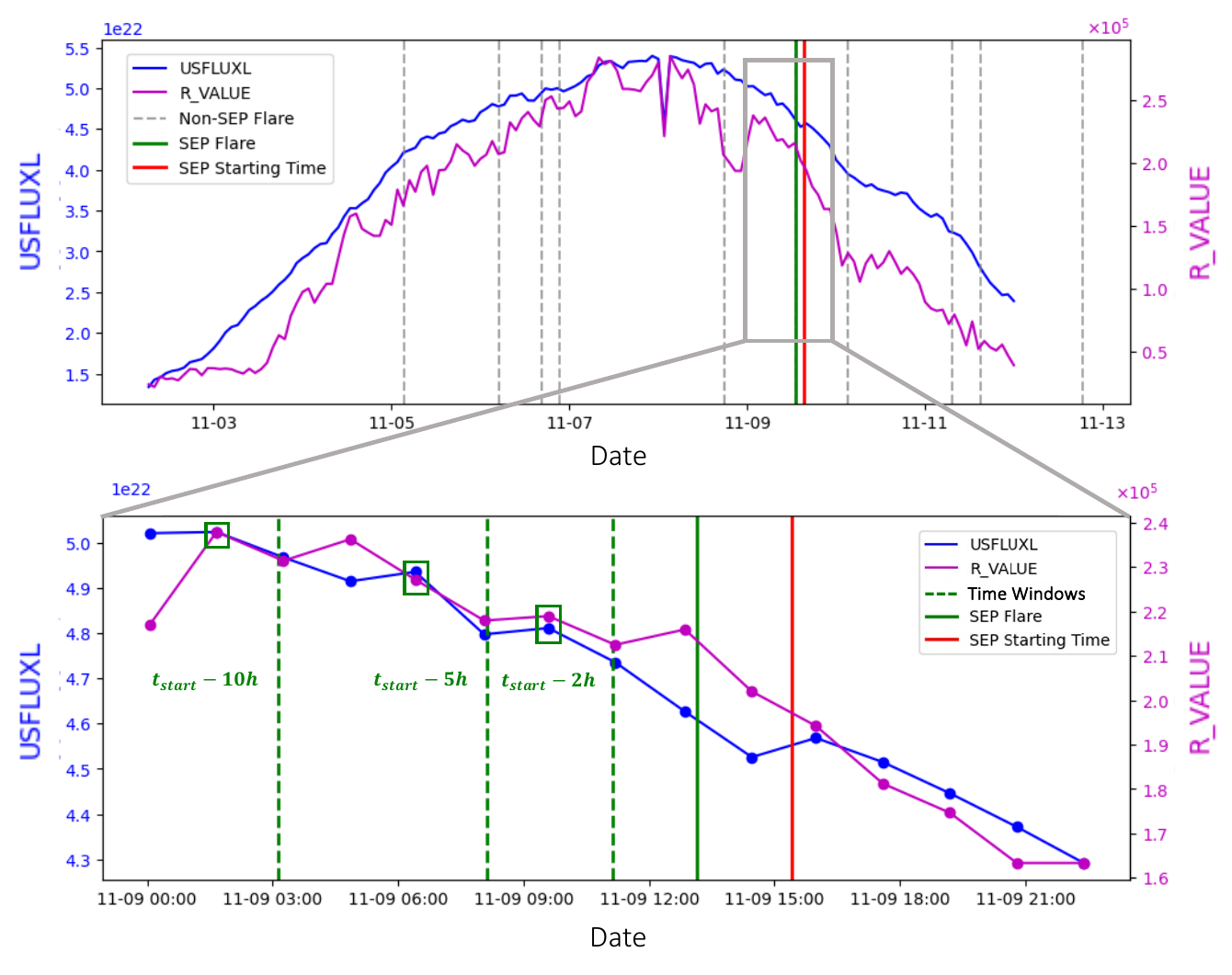}
\caption{Timelines of the total line-of-sight unsigned flux (USFLUXL in Maxwells) and the unsigned flux R near polarity inversion lines (R\_VALUE in Maxwells) for the total tracking period of AR$\,$10180 and the day (11/09/2002) during which a SEP occurred (bottom). Underlined (solid green) are the start time (\textit{t\_start} keyword) of the SEP producing (positive) flare of this AR and the data points that the matching algorithm selects for the ML-ready datasets. The process of selecting the data points in the green boxes is referred to as SMARP-SHARP Data Selection in Figure~\ref{fig:Schematic}.}
\label{fig:SHMARP_Picking}
\end{figure}

The procedure of preparing datasets (ML-Ready) for the training and validation of the machine learning models used in this research is illustrated in Figure~\ref{fig:SHMARP_Picking}. In blue and magenta are the USFLUXL and R\_VALUE timelines (Figure~\ref{fig:SHMARP_Picking}, top panel), which within a day show small fluctuations in value. Note that only 2 out of the 6 available keyword timelines (Figure~\ref{fig:Schematic}, SMARP-SHARP Dataset Timelines) are shown in Figure~\ref{fig:SHMARP_Picking}, annotated with the SEP starting time as defined by NOAA (red vertical line) and the starting times of the non-SEP-producing (negative) flares that occurred within AR 10180 (Figure~\ref{fig:SHMARP_Picking}, grey dashed lines). The moments of time 2, 5, and 10 hours before the start time of a SEP-producing (positive) flare are shown in green dashed lines (Figure~\ref{fig:SHMARP_Picking}, bottom only). These intervals are the different time windows this study produced results for (2-hour, 5-hour, 10-hour window). The forecasting window (or prediction window $t_p$, the time interval between green rectangles and lines in Figure~\ref{fig:SHMARP_Picking}) is defined as the selected time window plus the time to the first available data point. 

\begin{figure}[ht]
\centering
\includegraphics[width=\textwidth]{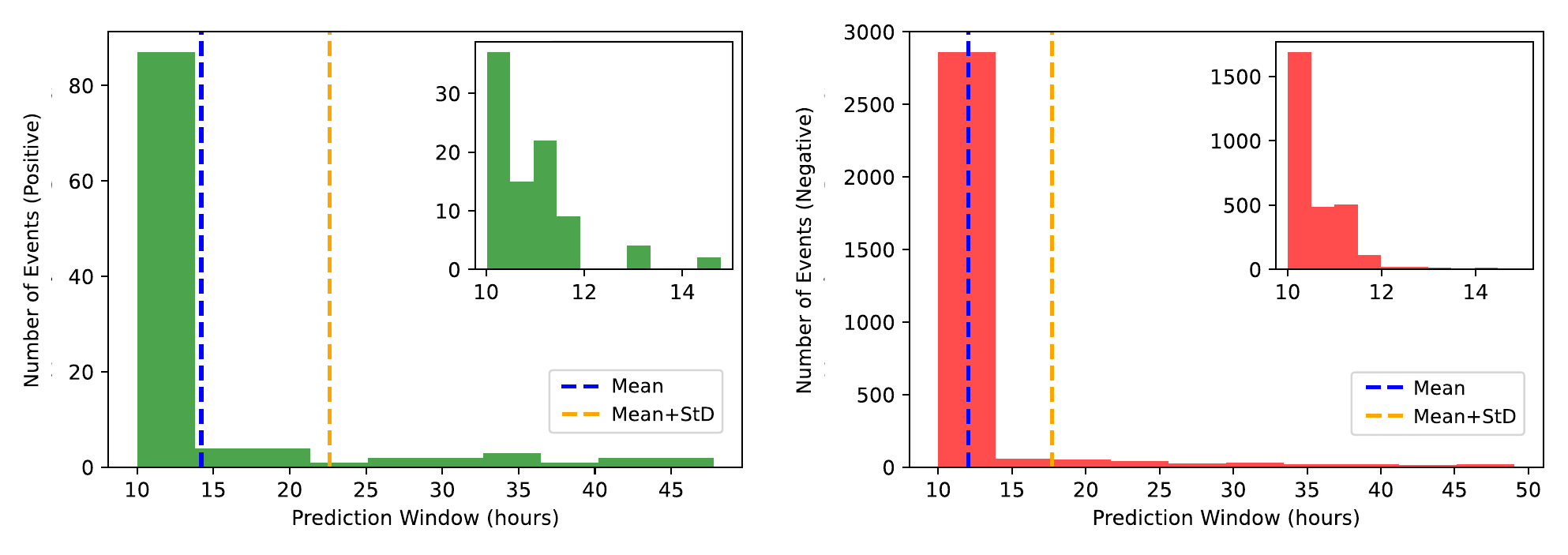}
\caption{Histograms of the minimum prediction windows variation ($t_p \geq t_{start} - 10$~ hours) for the 10-hour window ML-ready dataset. The inset histograms showcase the variation of prediction windows within the 1st bin. The data were split into 10 bins while the time between orange and blue dashed lines is equal to the prediction windows' ($t_p$) standard deviation (StD). The histograms in red are denoted the negative event values, and in green are the SEP-producing events.}
\label{fig:TimeWindow}
\end{figure}

The matching algorithm picks the first available SHARP-SMARP data point (Figure~\ref{fig:SHMARP_Picking}, bottom panel, green rectangles) before $t_{start} - x$ for $x \in [2,5,10]$ hours. This process is the same for all flares, regardless of whether they produced a SEP or not. These are the positive and negative data points (for three different time windows) that will comprise the ML-ready datasets discussed in this study. Every flare event has a unique prediction window $t_p \geq t_{start} - x $. In Figure~\ref{fig:TimeWindow} the histograms show the distribution of the different time windows ($t_{p}$) for the ML-ready dataset, for a case where the selected, in the matching algorithm, time window was 10 hours. Therefore, the dataset has a mean forecast window of 14.21 hours for the positive flares (orange dashed line, Fig.~\ref{fig:TimeWindow}, left panel) and 12.06 hours for the negative flares (orange line, right panel). Similar forecast windows were used in previous studies \citep{garcia2016prediction,anastasiadis2017predicting}, allowing for reliable comparison of results. 

Figure \ref{fig:TimeWindow} reveals flares (both positive and negative) that have a minimum prediction window of 48 hours, due to data gaps. This happens because of data gaps in the provided SHARP-SMARP dataset. Although these flares have undesirable forecast windows (data 2 days before the SEP occurrences is not as reliable for prediction), we still do not omit them as in this research, we are trying to preserve as many positive data points as possible (only 5 omitted as mentioned in Section \ref{sec:Matching}). 

\begin{figure}[ht]
\centering
\includegraphics[width=\textwidth]{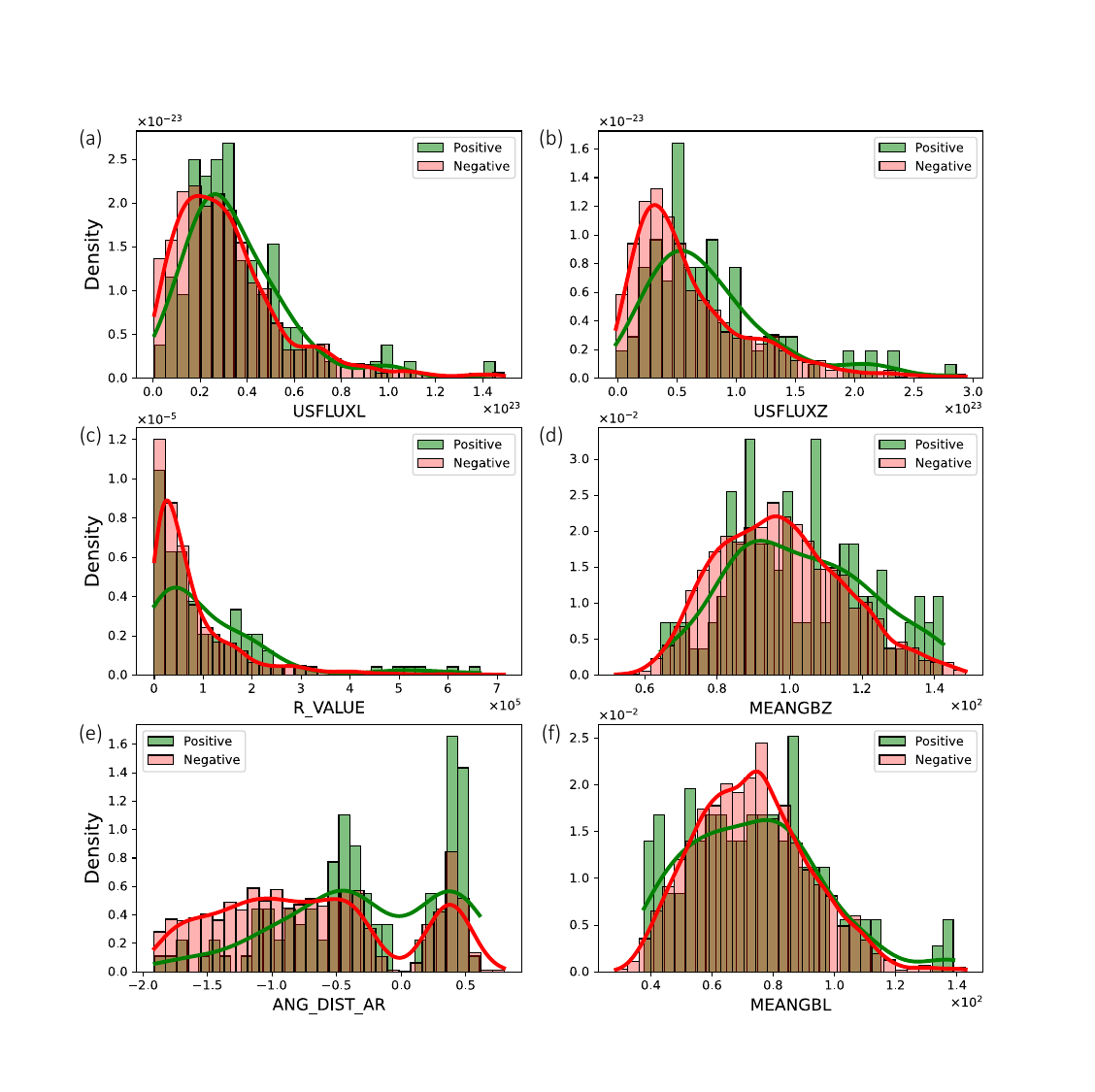}
\caption{Probability density histograms for the six SHARP-SMARP predictors (Table~\ref{tab:SHMARPKeywords}) for the 10-hour window ML-ready dataset for the positive (red) and negative (green) events. The red and green curves correspond to the fitted trend lines of the histogram bin values.}
\label{fig:PredictorsHistograms}
\end{figure}

To understand better the multi-dimensional problem that the ML model has to solve, we present in Figure~\ref{fig:PredictorsHistograms} density histograms of the six predictors (bold keywords, Table~\ref{tab:SHMARPKeywords}). The use of density instead of normal histogram values is essential when the populations of two different groups (positive and negative), are highly imbalanced like in our study. A strong SHARP-SMARP predictor would be one where the positive (green) and negative (red) event distributions have less overlap. The physical parameters that represent the mean value of the line-of-sight magnetic field gradient (MEANGBL, Figure~\ref{fig:PredictorsHistograms}f) offer less information about the eruption of a SEP. The angular distance of the AR and the magnetic footpoint of the Earth, along with the unsigned flux R near polarity inversion lines (R\_VALUE, Figure~\ref{fig:PredictorsHistograms}d), provide some information about the ability to distinguish between positive and negative flares and therefore the SEPs that can reach Earth. SEPs will unlikely start early in the evolution of the AR (or west of the magnetic footpoint of the earth as seen for AND\_DIST\_AR, Figure~\ref{fig:PredictorsHistograms}e). Two Gaussian distributions are visible at -0.4 and 0.4 rad from the magnetic footpoint. Therefore, some predictive capability can be observed in this single parameter, which, in combination with others, could distinguish whether a flare about to erupt can lead to a SEP or not. 

Similar to \cite{kasapis2022interpretable}, the predictors that carry most of the information beneficial to prediction tasks are those related to the unsigned magnetic flux (USFLUXL and USFLUXZ, Figure~\ref{fig:PredictorsHistograms} a, b). Although for both USFLUXL and USFLUXZ the overlap between positive and negative flares is significant (still less than the rest of the predictors), these predictors can carry meaningful information for an ML method to distinguish between the two populations. The difference between their distributions is noteworthy, with USFLUXZ showing better predictive capabilities. As expected, flares produced in AR areas where there is an increased value of magnetic flux, are more likely to produce an SEP. This is even more obvious for values of unsigned magnetic flux on the line-of-sight that are greater than 1.75 Maxwells (USFLUXZ $\geq 1.75$ Maxwells) as seen in panel b of Figure \ref{fig:PredictorsHistograms}. A similar behavior can be observed for large values ($\geq 120$ Gauss/Mm) of the line-of-sight field gradient (MEANGBZ, Figure~\ref{fig:PredictorsHistograms}d). 

% Add one sentence: that describes the splitting of flaring ARs in two groups (non-SEP producing and SEP-PRODUCING) - it might be good to reference Figure 2. Say what is first and second group if it is to use this terminology (do not use tho)
Following the SHMARP-SHARP data selection processes described in this section, we have created the first group of datasets for different time windows, which include information for all flares that occurred within the tracked ARs. We proceed with creating a second group of datasets (referred to as "Clean") by omitting the non-SEP-productive flares in the SEP-productive active regions as it is almost impossible for a machine learning method to distinguish between positive and negative data points that have such similar values. The difficulty here arises because an AR that hosts a SEP-producing flare, in most cases, also hosts multiple more non-SEP-producing flares that will have similar values with each other. Omitting the non-SEP-productive flares transforms the problem, as the forecasting now concerns the AR itself (and whether it is SEP-producing or not) rather than concerning the flare occurrences. In Figure~\ref{fig:SHMARP_Picking}, the data points related to the grey dashed lines ($t_{start}$ of the negative flares) are the ones omitted when training our SVM method on the "Clean" dataset. Tables \ref{tab:Statistics} and \ref{tab:TwoPredictors} show how we can present our ML models with an easier problem by omitting these noise-inducing flares. 

To quantify the histograms of Figure~\ref{fig:PredictorsHistograms}, the t-statistic and p-value of the positive and negative samples in our datasets are calculated and presented in Table \ref{tab:Statistics}. The first two columns under the 'Regular' category in the table represent the t-statistic and p-value for the dataset which includes all flares (not "Clean"). For instance, the keyword USFLUXL shows a t-statistic of 1.0418 and a p-value of 0.2976, indicating a moderate difference between the two groups with a relatively high probability of the result being due to chance. In contrast, ANG\_DIST with a t-statistic of 5.8973 and a p-value of $4.07 \times 10^{-9}$ signifies a more substantial difference with a much lower likelihood of the prediction result being random. This pattern is consistent across the keywords of all datasets. Keywords that show less overlap in the histograms, such as USFLUXZ, R\_VALUE, and ANG\_DIST, have higher t-statistic values, indicating a more pronounced difference between positive and negative samples. Conversely, keywords with more overlap, like MEANGBL, exhibit smaller t-statistic values and larger p-values, suggesting less distinction between the positive and negative groups. 

The increase of t-statistic values for USFLUXL, USFLUXZ, R\_VALUE, MEANGBZ, and especially MEANGBL suggests that cleaning the data has augmented the distinguishability of the means between positive and negative flares. This could lead to improved discrimination power when these variables are used as features in our prediction models. However, it is important to note the very slight increase in the p-value for ANG\_DIST upon cleaning. This underlines the boundaries of predictive power within the SHARP-SMARP dataset, even when the prediction problem becomes easier.

\begin{table}[htbp]
\centering
\begin{tabular}{lccccc}
\hline
Keyword & \multicolumn{2}{c}{Regular} & \multicolumn{2}{c}{Clean} & Difference \\
\cline{2-5}
& t-statistic & p-value & t-statistic & p-value & t-statistic / p-value \\
\hline
USFLUXL & 1.0418 & 0.2976 & 2.3254 & 0.0201 & +1.2836 / -0.2775 \\
USFLUXZ & 3.2500 & 0.0012 & 5.1861 & $2.31 \times 10^{-7}$ & +1.9361 / -0.0009 \\
R\_VALUE & 2.9673 & 0.0030 & 6.4370 & $1.44 \times 10^{-10}$ & +3.4697 / -0.0029 \\
MEANGBZ & 2.9293 & 0.0034 & 4.2548 & $2.17 \times 10^{-5}$ & +1.3255 / -0.0032 \\
ANG\_DIST & 5.8973 & $4.07 \times 10^{-9}$ & 5.9810 & $2.51 \times 10^{-9}$ & +0.0837 / +0.0005 \\
MEANGBL & -0.3880 & 0.6980 & 0.8734 & 0.3825 & +1.2614 / -0.3155 \\
\hline
\end{tabular}
\caption{The t-statistic and p-value coefficients for the six keywords of the regular and clean datasets. The last two columns include the difference in the t-statistic and p-value between the regular and clean datasets.}
\label{tab:Statistics}
\end{table}

The flare intensity statistic values are indicative of the predictive power the flare information holds. The flare intensity t-statistic value is 16.68, almost three times greater than the t-statistic for ANG\_DIST (5.8973). Similar observations can be done for the p-values proving the consistency of these results. For the physical parameter that the two datasets (flare and SHARP-SMARP) share in common, the distance to the magnetic footpoint of the Earth, the t-statistic of 6.66 for \textit{ang\_dist} (Table~\ref{tab:FlareKeywords}) is higher than that of ANG\_DIST, underlining again the superiority of the flare data in predictive power. The only difference between ANG\_DIST and \textit{ang\_dist} is that in the first case, the values are calculated using the AR midpoint coordinates, whereas for the flares they are taken from the exact point the flare erupted. The slight increase in t-statistic and decrease in p-value prove that more exact coordinate values for the SEP location would have beneficial effects in an ML model. Given the concrete knowledge these statistic tests provide, in the following Sections we will be using the SMARP-SHARP dataset predictors to forecast SEP events at least half a day (14 hours) before they are observed.

%%%% FEB 9 revision w irina %%%%%%%%

%%%%%%% NEW SECTION %%%%%%% 

\section{Machine Learning Models and Methodology} \label{sec:ML}

To evaluate the predictive capabilities of the SMARP-SHARP dataset for SEP prediction, we use the Support Vector Machine (SVM) and Linear Regression models available in the Scikit-Learn\footnote{\url{https://scikit-learn.org/stable/modules/generated/sklearn.svm.SVC.html}} Python library (\textit{sklearn}). Support Vector Machines \citep{hearst1998support,steinwart2008support} are a type of supervised machine learning model used for classification or regression tasks. SVMs aim to find the optimal hyperplane that best separates data points of different classes in a high-dimensional space. They are particularly effective in high-dimensional spaces such as ours, as we deal with six  SMARP-SHARP predictors (six dimensions) and when the classes are linearly separable. Since there is no certainty about the linear separability of our data (Figure \ref{fig:PredictorsHistograms}), especially in their original feature space, in this research we explore different SVM kernels. The kernel trick involves mapping the data into a higher-dimensional space where it becomes linearly separable, or more separable than in the original space. The SVM kernels used in this research are: a) the Polynomial Kernel (\textit{poly}: finds a polynomial of the given degree to separate the data), b) the Radial Basis Function or Gaussian Kernel (RBF: creates a landscape where data points that are close in the original space are at the peak, and those further away are down the slope) and the Sigmoid Kernel (\textit{sigmoid}: maps the similarity between data points into values between -1 and 1). 

The second group of ML models is regression models \citep{fahrmeir2013regression}, which are a set of statistical methods for estimating the relationships among variables. They are used to predict a continuous outcome variable (dependent variable) based on one or more predictors (independent variables). The most common type of regression analysis is linear \citep{weisberg2005applied}, where a line of best fit is determined, but other types of models include logistic \citep{nick2007logistic}, polynomial \citep{heiberger2009polynomial} and Ridge \citep{marquardt1975ridge} regression. While linear regression is traditionally used for predicting continuous numerical values, there are variants of linear models suitable for non-continuous or categorical data such as the ones discussed in this work (SEP/no-SEP). Logistic regression used in this research, for instance, is designed for binary classification tasks that fit our positive/negative (SEP/No-SEP) prediction problem.

These ML models are dependent on a set of hyperparameters, which play a crucial role in determining the performance of the prediction capability. While model parameters are learned during training, hyperparameters are external to the model, they often control its overall behavior. The choice of the kernel (linear, polynomial, RBF, etc.) and their respective parameters, like the degree for a polynomial kernel or gamma for RBF, are all considered as hyperparameters. Similarly, in logistic regression, the regularization strength and type (L1 or L2) are hyperparameters that can influence the model's performance. Manually selecting these hyperparameters can be suboptimal; we, therefore, employ tools like GridSearchCV\footnote{\url{https://scikit-learn.org/stable/modules/generated/sklearn.model_selection.GridSearchCV.html}} from Scikit-Learn, which offers an automated and systematic approach to hyperparameter tuning. By searching through specified hyperparameter combinations and cross-validating, the GridSearchCV algorithm ensures the selection of hyperparameters that yield the best model performance. However, it's worth acknowledging the computational expense of such a brute-force approach, especially when considering extensive hyperparameter spaces or sizable datasets. In this study, we train our models using 6 dimensional feature vectors in the amount of low thousands therefore, The computational expense is in the order of minutes. In the following section, SEP prediction results are produced by six ML methods: Linear, RBF, Polynomial, and Sigmoid SVMs, along with Logistic and Ridge Regression. 

To assess the performance of each machine learning model, a suite of evaluation metrics was chosen (Table \ref{tab:TwoPredictors}). Accuracy (ACC) measures the proportion of correct predictions made by the model relative to the total number of predictions. For ACC, a score of 1 indicates perfect prediction, while a score of 0.5 suggests no better than random guessing. True Skill Statistic (TSS) and Heidke Skill Score (HSS) both account for the skill of the model in distinguishing between the classes, factoring in both false positives and false negatives. For both TSS and HSS, a score of 1 indicates perfect skill, 0 indicates no better than random prediction, and negative values indicate inverse or contrary predictions. False Alarm Rate (FAR) measures the proportion of negative instances that were incorrectly classified as positive; a lower FAR is more desirable, with 0 being the ideal score. Lastly, the F1 Score provides a balance between precision and recall. An F1 score closer to 1 indicates better balance and performance, while a score closer to 0 suggests poor performance. These metrics provide a comprehensive evaluation framework, ensuring the model's performance is assessed from multiple vantage points.

To robustly evaluate the predictive ability of the various predictors discussed in Section \ref{sec:Statistics} and to quantify model uncertainty, each model was trained and validated across 100 distinct runs. Therefore, the results reported in this work reflect the mean value and standard deviation of 100 distinct ML training and testing runs that used the same hyperparameters but different initializations. A one-to-ten (0.1) train-to-test ratio is used for the validation of every individual run. Our ML-ready datasets include 110 positive events, therefore 99 of them are used for training and 11 for testing. The methodology employed for training the aforementioned machine learning models was designed to account for the severe class imbalance in the dataset. There are 3,356 negative samples (non-SEP-producing flares) in contrast to only 110 positive samples (SEP-producing flares), resulting in an imbalance ratio of 1/30. Two types of training and validation schemes are used in our analysis: one where the positive and negative datasets are balanced (referred to as Balanced) and one where the on-to-thirty imbalance is retained (Imbalance). 

In the Balanced case, every run utilizes all positive samples (110) and an equal number of negative samples, randomly chosen out of the total population (3,356). Therefore the 110 chosen samples are different in each one of the 100 runs, ensuring a fair representation of the entire population. This experimental configuration (referred to as Balanced in Table \ref{tab:TwoPredictors}) was intentionally chosen to discern the inherent predictive capability of the predictors in a balanced, experimental scenario. In this case testing is performed using the same amount of negative and positive samples. In the second case (labeled Imbalanced in Table \ref{tab:TwoPredictors}), rather than randomly picking an equal amount of negative samples, we used the entire dataset (retains the 1/30 imbalance) but added a weighting factor which during training allows for more attention to the minority class (positive). Here, the number of negative samples reserved for testing is chosen to be 30 times greater (3300 negative testing samples) than that of positive.

%%%%%%% NEW SECTION %%%%%%% 

\section{Results} \label{sec:Results}

In the pursuit to identify the most effective machine learning model for SEP prediction, a series of experiments were conducted with different combinations of the six SMARP-SHARP predictors (Figure~\ref{fig:PredictorsHistograms}). We performed tests with all possible predictor combinations. In this paper, we present the five best combinations. Across these tests, the performance of the models in terms of accuracy remained within a relatively narrow band, from \(0.53 \pm 0.10\) for the USFLUXL, MEANGB, MEANGBZ combination (Table~\ref{tab:PredictorsResults}), to \(0.67 \pm 0.11\) for the R\_VALUE, ANG\_DIST predictors. It is noteworthy that although the R\_VALUE has a lower t-statistic value (and higher p-value) than USFLUXZ, when used along with the lowest t-statistic predictor in Table~\ref{tab:Statistics} (ANG\_DIST), produces slightly better results than if the ANG\_DIST was combined with the USFLUXZ. This pair consistently achieved the highest accuracy across all models, reaching \(0.67 \pm 0.09\) when training an SVM model that uses the Sigmoid kernel. 

\begin{table}[]
\centering
\begin{tabular}{llcccclcc}
&  & \multicolumn{4}{c}{SVM}  &  & \multicolumn{2}{c}{Regression} \\ \cline{3-6} \cline{8-9} 
SMARP-SHARP Predictors  &  & Linear  & RBF   & Polynomial  & Sigmoid  &  & Logistic  & Ridge  \\ \cline{1-1} \cline{3-6} \cline{8-9} 
ALL PREDICTORS &  & $0.64 \pm 0.10$ & $0.62 \pm 0.10$ & $0.64 \pm 0.10$  & $0.62 \pm 0.10$ &  & $0.65 \pm 0.09$ & $0.65 \pm 0.12$ \\
USFLUXZ, R\_VALUE, ANG\_DIST &  &  $0.66 \pm 0.09$  & $0.65 \pm 0.10$  & $0.66 \pm 0.10$ &  $0.64 \pm 0.10$ &  & $0.65 \pm 0.10$ & $0.65 \pm 0.09$  \\
USFLUXL, MEANGB, MEANGBZ &  & $0.55 \pm 0.10$ & $0.55 \pm 0.10$ & $0.54 \pm 0.10$ & $0.53 \pm 0.10$ &  & $0.56 \pm 0.10$ & $0.56 \pm 0.11$ \\
R\_VALUE, ANG\_DIST  &  & $0.67 \pm 0.10$ & $0.66 \pm 0.10$ & $0.67 \pm 0.11$ & $0.67 \pm 0.09$ &  & $0.66 \pm 0.10$ & $0.65 \pm 0.10$ \\
USFLUXZ, ANG\_DIST  &  & $0.66 \pm 0.10$ & $0.64 \pm 0.11$ & $0.64 \pm 0.10$ & $0.66 \pm 0.10$ &  & $0.66 \pm 0.11$ & $0.65 \pm 0.10$ \\
USFLUXZ, R\_VALUE &  & $0.54 \pm 0.11$ & $0.58 \pm 0.11$ & $0.54 \pm 0.08$ & $0.58 \pm 0.10$ &  & $0.57 \pm 0.09$ & $0.55 \pm 0.11$       
\end{tabular}
\caption{Accuracy (ACC) values for SVM and Regression models when trained on six combinations of the SMARP-SHARP predictors for runs that include all flares (Regular Dataset in Table~\ref{tab:TwoPredictors}) and a 10-hour time window.}
\label{tab:PredictorsResults}
\end{table}

Surprisingly, incorporating more predictors did not translate to enhanced performance. For instance, using all available predictors yielded accuracies such as \(0.64 \pm 0.10\) (0.03 decrease compared to examples with the best performance, Table~\ref{tab:PredictorsResults}) for the Linear SVM model, which is not superior to using just two predictors, as seen with R\_VALUE and ANG\_DIST. This can be attributed to the inclusion of low-quality features that may introduce noise to the ML training rather than providing valuable information. Observing the SVM model results (Table~\ref{tab:PredictorsResults}), it is evident that the Linear, RBF, and Polynomial kernels consistently demonstrate similar performance across most predictor combinations. Similarly, the regression models, both Logistic and Ridge, show competitive performance compared to the SVM models. Their accuracy values are in line with the top-performing SVM models, and the consistency in their performance is evident from the relatively small standard deviations. 

\begin{figure}[ht]
\centering
\includegraphics[width=0.95\textwidth]{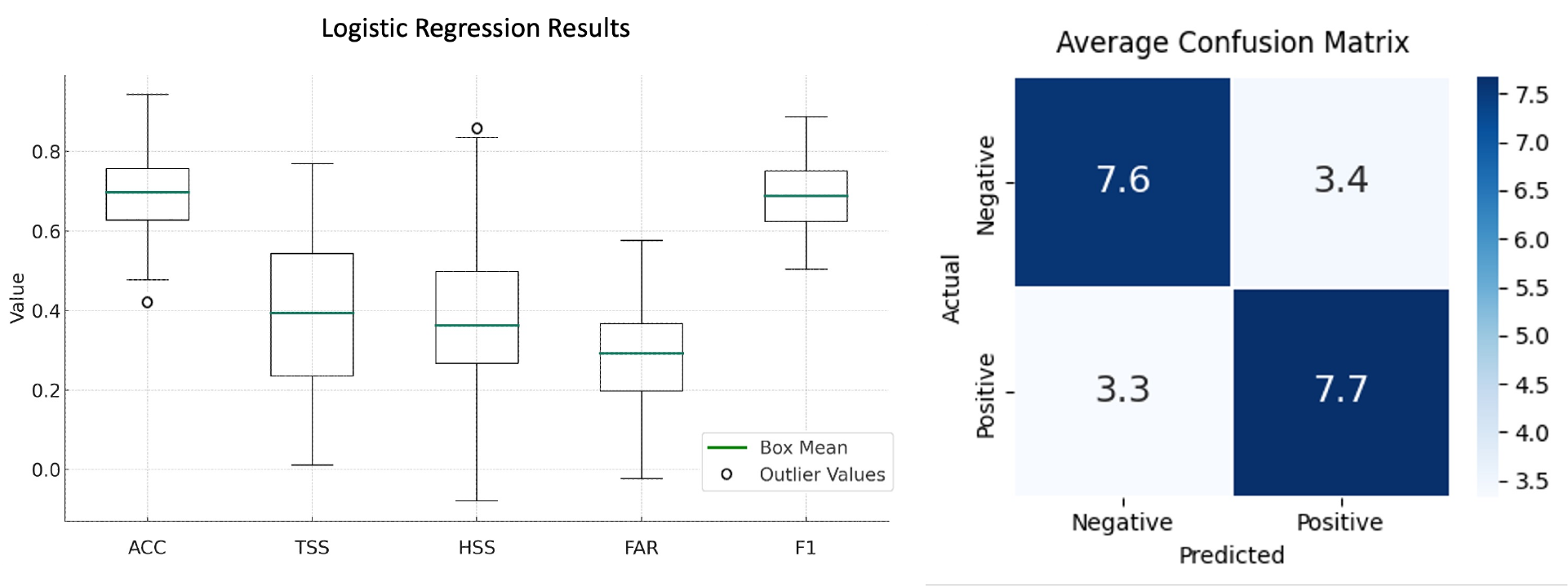}
\caption{The distribution of 100 score values of  ACC, TSS, HSS FAR, and F1 (left panel). 
Each metric is calculated using the entries of the confusion matrix produced in each run, the average over 100 runs (right panel). The values were obtained using the R\_VALUE and the ANG\_DIST on a Logistic Regression model and constitute the best SEP prediction the SMARP-SHARP data can achieve. The box range shows the interquartile range, the green line inside is the median value, the whiskers show the results range, and the circles show two outlier values. The samples are obtained using the Clean dataset, where non-SEP-producing flares in SEP-producing ARs are omitted and used on a balanced training method where positive and negative training and evaluation datasets are of equal size.}
\label{fig:LogisticRegression}
\end{figure}

Note that for every value tabulated in Table~\ref{tab:PredictorsResults}, 100 different runs are performed using an equal amount (110) of positive and negative data points (Balanced), the same hyperparameters, different model initialization, and a different set of randomly picked negative data points. The training of each model is performed using 90\% of the 220 ($2 \times 110$) available data points while 10\% of the data points (22) are unseen during training and reserved for evaluation (198 training vectors, 99 positive and 99 negative and 22 evaluation vectors, 11 positive and 11 negative). Similar training and testing configurations are used to produce the results of Table~\ref{tab:TwoPredictors}, but for the two different training processes (Balanced and Imbalanced) and the two different groups of datasets (Regular and Clean) discussed in previous sections. The training and dataset selection are the factors that define whether the setting of a test is closer to simulating an operational setting (labeled as Operational in Table~\ref{tab:TwoPredictors}) or the setup is more experimental (Semi-Operational and Experimental). The Experimental results are closely examined in Figure~\ref{fig:LogisticRegression}.

\begin{table}[h]
\centering
\begin{tabular}{lccccccccc}
Setting & Training & Dataset & Model  &  & ACC & TSS & HSS & FAR & F1 \\ \hline
Operational & Imbalance & Regular & SVM Linear &  & $0.56 \pm 0.04$ & $0.32 \pm 0.10$ & $0.05 \pm 0.02$ & $0.24 \pm 0.10$ & $0.71 \pm 0.04$ \\ 
{\footnotesize \cite{kasapis2022interpretable}} & Imbalance & Regular & SVM Poly &  & $0.52 \pm 0.05$ & $0.01 \pm 0.01$ & $0.01 \pm 0.01$  & $0.98 \pm 0.05$ & $0.58 \pm 0.06$ \\
Semi-Operational & Balanced & Regular & SVM Linear &  & $0.66 \pm 0.10$ & $0.33 \pm 0.19$ & $0.33 \pm 0.19$ & $0.26 \pm 0.14$ & $0.63 \pm 0.12$  \\
{\footnotesize \cite{kasapis2022interpretable}} & Balanced & Regular & SVM RBF &  & $0.65 \pm 0.13$ & $0.33 \pm 0.28$ & $0.31 \pm 0.26$ & $0.31 \pm 0.10$ & $0.75 \pm 0.04$ \\
Semi-Operational & Imbalance & Clean & SVM Linear &  & $0.67 \pm 0.04$ & $0.35 \pm 0.13$ & $0.08 \pm 0.03$ & $0.32 \pm 0.12$ & $0.79 \pm 0.03$ \\
Experimental & Balanced & Clean & SVM Linear &  & $0.69 \pm 0.09$ & $0.38 \pm 0.19$ & $0.38 \pm 0.19$ & $0.26 \pm 0.14$  & $0.67 \pm 0.11$ \\
Experimental & Balanced & Clean & Logistic Reg &  & $0.70 \pm 0.09$ & $0.39 \pm 0.19$ & $0.37 \pm 0.19$ & $0.30 \pm 0.14$ & $0.69 \pm 0.10$ 
\end{tabular}
\caption{Results for a variety of ML models and training configurations when using the strongest predictors couple, the R\_VALUE and the ANG\_DIST.}
\label{tab:TwoPredictors}
\end{table}

To comprehend the results in a setting that better addresses the operational needs of a SEP prediction apparatus, we adopted a different approach where each ML model is trained on the entire dataset, and a weighting factor, \textit{class\_weight} (parameter in the Scikit-Learn python library) is introduced to account for the class imbalance problem. For the "Imbalance" models (Table \ref{tab:TwoPredictors}), the weight for the negative class was set to be the ratio (1/30) of positive to negative samples in the training set. This approach allows the model to emphasize the minority class during training, compensating for the lack of negative samples. The results obtained from this methodology, as presented in Table \ref{tab:TwoPredictors}, reveal that the accuracy cannot remain competitive compared to the balanced and therefore experimental setting. More specifically, a significant drop in evaluation metrics such as ACC, TSS, HSS, and F1 is observed, with the largest being a $0.10 \pm 0.09$ decrease in accuracy. Despite this fact, the F1 score for the operational ML method imbalanced for the entire dataset is higher than when we train on an experimental balanced scheme (last row, Table~\ref{tab:TwoPredictors}). This suggests that, in general, although the operational model's overall capability to correctly predict both positive and negative classes has decreased to almost a random pick, the model makes fewer false negatives (increased recall), and the accuracy of positive predictions (precision) remains high, even if the overall accuracy has dropped. Such trade-offs are often seen in ML binary classification.

Another two types of learning schemes are employed, one in which all available flares are included (Original Dataset) and one where all non-SEP-producing flares that have occurred in a SEP-producing active region (Figure~\ref{fig:Schematic}, Clean Datasets) have been omitted. This is done because, as demonstrated in Figure \ref{fig:SHMARP_Picking}, it is almost impossible for an ML model trained on low-dimensional data to distinguish between a positive and a negative flare sample that has occurred almost at the same time in very close proximity. As expected, for both the balanced and imbalanced runs in Table \ref{tab:TwoPredictors}, the Clean dataset runs exhibit higher predictive performances in the majority of the metrics analyzed. A slight increase in accuracy ($+0.3\pm 0.1$) is observed when training our models using a balanced dataset, whereas a significant increase ($+0.11 \pm 0.1$) can be observed when training with the Clean dataset a Linear SVM with imbalance. This is notable because it shows that even when training with imbalance, an ML method can predict whether an AR (instead of a flare) will be SEP-productive or not.

Table~\ref{tab:TwoPredictors} shows that both the Operational and the Semi-Operational configuration results are comparable to the corresponding results by \cite{kasapis2022interpretable} for similar settings. In an operational setting, an SVM model performs better (ACC increase of $0.04\pm0.1$ and F1 increase of $0.13\pm0.1$) when trained on data from two Solar Cycles compared to if it was trained only using the SMARP parameters. Regardless of this increase, our efforts to simulate an operational setting by conserving the inherited imbalance led to predictions that are marginally better than a random guess (Table~\ref{tab:TwoPredictors}, Operational ACC=$0.56\pm0.04$). When shifting to a Semi-Operational setting, the increase in ACC, TSS and HSS scores for an SVM model (Table~\ref{tab:TwoPredictors}, third row) trained on almost double the amount of data, compared to the results of our previous work (fourth row), is within the margin of error, therefore insignificant. In this study, the FAR has decreased by $0.05\pm0.1$, showing that the increase in positive events allows the SVM models to reduce the number of false alarms. The highest accuracy achieved in \cite{kasapis2022interpretable}, for an Experimental setting, is $0.72 \pm 0.12$ on a 3rd-degree polynomial SVM when training on USFLUXL and ARDIST (equivalent of ANG\_DIST in this work)- results that could not be reached using the equivalent SMARP-SHARP dataset predictors. In this work, the R\_VALUE and ANG\_DIST combination of parameters is the one that showed the best performance. This shows that the recomputed R\_VALUE in the dataset provided by \citep{kosovich2024time} provides meaningful information to ML models in regard to SEP production.

\begin{figure}[ht]
\centering
\includegraphics[width=0.75\textwidth]{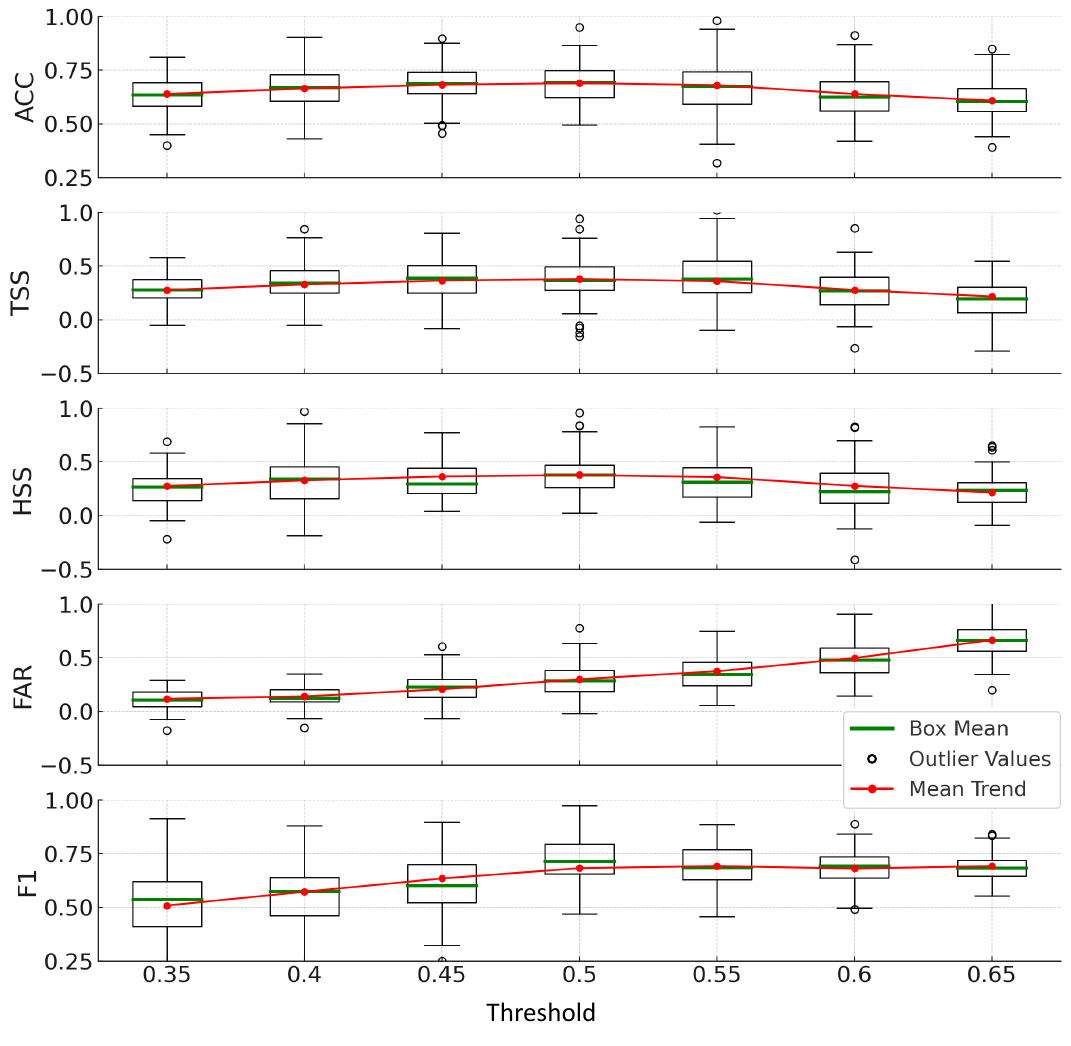}
\caption{Box plots of the distribution of 100 score values of  ACC, TSS, HSS FAR, and F1, for seven decision thresholds $\tau$. The values were obtained using the R\_VALUE and the ANG\_DIST to train an Experimental Logistic Regression model. The box range shows the interquartile range, the green line inside it shows the median value, the whiskers show the results range, and the circles show two outlier values. These results correspond to the seventh row in Table~\ref{tab:TwoPredictors} for $\tau = 0.5$.}
\label{fig:LogisticRegression2}
\end{figure}

The results procured from the best-performing, in terms of accuracy, run of the Experimental Logistic Regression model in Table~\ref{tab:TwoPredictors} (seventh row), trained on a balanced positive and negative setting, offer some interesting insights. The model showcases an accuracy of $0.70 \pm 0.093$ over 100 runs, and the TSS and HSS both register an average of $0.386 \pm 0.187$. The FAR of $0.303 \pm 0.138$ indicates a moderate rate of false alarms, while the F1 score is $0.688 \pm 0.099$, representing a harmonization of precision and recall. In this experimental case, where the problem is reduced to predicting whether an active region will produce a SEP, the SMARP-SHARP dataset shows a notable forecasting capability, comparable to other studies \citep{garcia2016prediction,nunez2011predicting,kasapis2022interpretable}. 

To examine the models' performance, it's essential to consider the combination of different metrics. Figure \ref{fig:LogisticRegression2} shows that adjusting the threshold in Logistic Regression impacts the different performance metrics. By default, many machine learning models, including logistic regression, employ a threshold of 0.5, whereby probabilities greater than this value are designated as the positive class, and those below are designated as the negative class. Lowering the threshold from $0.65$ to $0.35$ generally reduces the false alarm rate, from $0.66$ ($\tau = 0.65$) to $0.118$ ($\tau=0.35$), indicating fewer false positives. However, this benefit comes with a trade-off, as accuracy drops to $0.62$ at $\tau = 0.65$. The F1 Score, which represents the balance between precision and recall, is maximum when $\tau=0.55$, suggesting this may be the best threshold for a harmonized performance. Selecting the right threshold involves balancing these trade-offs, aiming for minimal false alarms (lower FAR) or higher overall accuracy and skill (higher ACC, TSS, and HSS).

%%%%%%% NEW SECTION %%%%%%% 

\section{Conclusion} \label{sec:Conclusion}

To refine SEP event prediction, this work utilizes a new dataset that combines the SMARP and SHARP parameters covering Solar Cycles 23 and 24 \citep{kosovich2024time}. The SMARP-SHARP dataset provides physical parameters for 110 active regions that produced SEP events. Due to differences in their problem formulation, it is difficult to make fair comparisons between works on SEP prediction in space weather literature. Despite this fact, in this work, a fair comparison is possible with \cite{kasapis2022interpretable} as both sets of experiments use the exact same experimental setup, with the only difference being the amount of data (two Solar Cycles instead of one). 

A set of ML models was trained on a number of datasets that differ in the number of predictors used (varying from 2 to 6), their type (USFLUX, R\_VALUE, MEANGBL, USFLUXZ, MEANGBZ, and ANG\_DIST), and the time window (2, 5 and 10 hours) their selected data offer. Our investigation reveals that when the SMARP-SHARP dataset-based ML model faces a problem of detecting whether an AR will produce a SEP, regardless of how many flares occurred within it, the resulting accuracy of $0.70\pm0.09$ is comparable to the results of the author's previous study. When encountering the problem of predicting whether a flare produces a SEP or not, the predictive power of the dataset diminishes. Although the results for a SMARP-SHARP dataset-based ML model appear better than if only SMARPs were used, for an operational setting, the accuracy results are only slightly better (ACC = $0.56\pm0.04$) than a random pick. 

Our results' similarities with those in previous studies, underscores the inherent complexity of SEP prediction; even with increased data (double in volume), the ceiling of accuracy remains consistent. Interestingly, our results indicate a modest increase of $0.04 \pm 0.05$ in ACC but significant improvements for the TSS, HSS, FAR, and F1 metrics in the operational context, where data imbalance is introduced. The SMARP-SHARP dataset contributes positively here, yet the overall accuracy achieved still denotes a considerable margin for enhancement. This suggests that the predictive capability of the data, while evident, lacks the reliability required for confident operational forecasting. The low dimensionality of the SMARP-SHARP data hinders our models' ability to distinguish between SEP-producing and non-SEP-producing flares that occur within the same AR. This is also evident when testing different prediction windows, where the SMARP-SHARP values do not change enough between the 2, 5, and 10-hour window, and therefore, the trained models did not produce noticeably different results. Consequently, we acknowledge the limitations of low-dimensional datasets and advocate for the exploration of more sophisticated, high-dimensional methodologies that may understand better and unravel the intricate patterns of SEP events.

Lastly, our study shed light on the relevance of the SMARP-SHARP dataset physical parameters (keywords) to SEP events. This study verifies that the total line-of-site unsigned magnetic flux, the distance of the flare location to the magnetic foot-point of the Earth, and the unsigned flux R near the polarity inversion lines are physical quantities that relate to the production of SEP events. It is recommended that future studies use them for the prediction of SEP events. The importance of magnetic connectivity between the flare location on the solar disc and the Earth is also shown for a SEP to be detected. 

\section*{Acknowledgments}

This work is partially supported by the NASA Heliophysics Supporting Research Grant, the NASA grants 80NSSC19K0630 and 80NSSC20K0302, and NSF grants 1639683 and 1916509. The code for this project is available at \url{https://github.com/skasapis/SEP\_Pred\_SMARP-SHARP}. We acknowledge the SOLSTICE Team at the University of Michigan, Department of Climate and Space Sciences and Engineering for their contributions to the methodology of this work.

%Acknowledgement by Sasha.

\bibliography{sample631}{}
\bibliographystyle{aasjournal}

\end{document}